\documentclass[aps,prl,twocolumn]{revtex4}
\usepackage{amsmath,bm,epsfig}

\def\Fbox#1{\vskip1ex\hbox to 8.5cm{\hfil\fboxsep0.3cm\fbox{%
  \parbox{8.0cm}{#1}}\hfil}\vskip1ex\noindent}  

\let \= \equiv \let\*\cdot \let\~\widetilde \let\^\widehat \let\-\overline

\def\<{\left\langle}    \def\>{\right\rangle}
\def\({\left(}          \def\){\right)}
 \def \[ {\left [} \def \] {\right ]}

\begin{document}
\title{Quantitative Theory of a Relaxation Function in a Glass-Forming System}
\author{ Edan Lerner and Itamar Procaccia}
\affiliation{Department of Chemical Physics, The Weizmann
Institute of Science, Rehovot 76100, Israel }
\date{\today}
\begin{abstract}
We present a quantitative theory for a relaxation function in a simple glass-forming model (binary mixture
of particles with different interaction parameters). It is shown that the slowing down is caused by the competition between locally favored regions (clusters) which are long lived but each of which relaxes as a simple function of time. Without the clusters the relaxation of the background is simply determined by one typical length which we deduce from an elementary statistical mechanical argument. The total relaxation function (which depends on time in a nontrivial manner) is quantitatively determined as a weighted sum over the clusters and the background. The  `fragility' in this system can be understood quantitatively since it is determined by the temperature dependence
of the number fractions of the locally favored regions.   

\end{abstract}
 \maketitle
Two fundamental riddles in glass-forming systems \cite{96EAN,01Don} are (i) what determines the spectacular
slowing down of the relaxation to equilibrium when the temperature is lowered through a relatively short interval, and (ii) how to predict theoretically the functional forms of various relaxation functions. In practice one usually fits  the data to phenomenological relaxation functions (e.g. the Kohlrausch-Williams-Watt KKW law) without any theoretical justification. Here we employ a classical example of glass-formation, i.e. a binary mixture of particles with different interaction diameters, to demonstrate unequivocally that the slowing down is due to the creation of clusters of local order; these are mechanically stable and slow to relax. We present a {\em quantitative} computation of a (functionally non-trivial) and strongly temperature-dependent relaxation function by presenting it as a weighted sum of  cluster contributions each of which decays as a simple relaxation function. 

\noindent
{\bf The model} discussed here is the classical example \cite{89DAY,99PH} of a glass-forming binary mixture of $N$ particles in a 2-dimensional domain of area $A$, interacting via a soft $1/r^{12}$ repulsion with a `diameter' ratio of 1.4. More or less related models can be found in \cite{FA1984,davisonsherrington2000,sh2002,schliecker2002,Jstatmech2007,benarous2006}.  We refer the reader to the extensive work done on this system \cite{89DAY,99PH,07ABHIMPS,07HIMPS,07IMPS}. The sum-up of this work is that the model
is a {\em bona fide} glass-forming liquid meeting all the criteria of a
glass transition. In short, the system consists of an
equimolar mixture of two types of particles, ``large"  with `diameter' $\sigma_2=1.4$ and ``small" with 'diameter` $\sigma_1=1$, respectively, but with the same mass $m$. The three pairwise
additive interactions are given by the purely repulsive soft-core potentials
\begin{equation}
u_{ab} =\epsilon \left(\frac{\sigma_{ab}}{r}\right)^{12} \ , \quad a,b=1,2 \ ,
\label{potential}
\end{equation}
where $\sigma_{aa}=\sigma_a$ and $\sigma_{ab}= (\sigma_a+\sigma_b)/2$. The
cutoff radii of
the interaction are set at $4.5\sigma_{ab}$. The units of mass, length, time
and temperature are $m$, $\sigma_1$, $\tau=\sigma_1\sqrt{m/\epsilon}$ and
$T=\epsilon/k_B$, respectively, with $k_B$ being
Boltzmann's constant. The results presented below are extracted from molecular dynamics 
simulations using 30 independent systems of 4096 particles each, using the Nose-Poincare-Andersen
thermostat \cite{00SL}. We employ periodic boundary conditions on the torus.

\begin{figure}
~\hskip -0.6 cm
\includegraphics[width=0.55\textwidth]{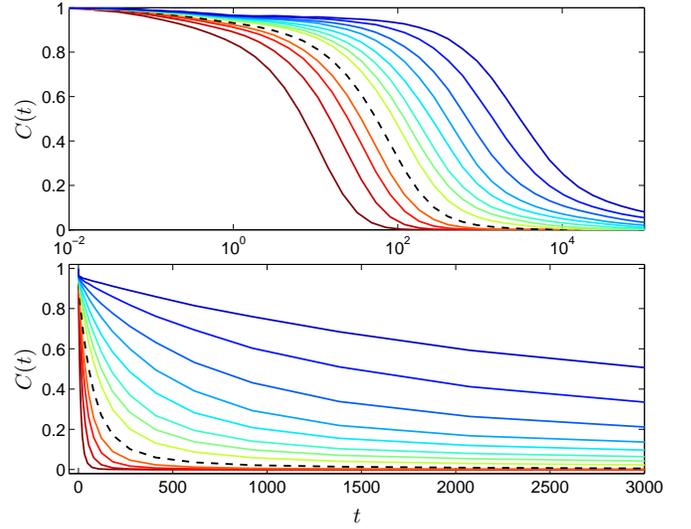}
\caption{(Color online). The relaxation function $C(t)$ as a function of logarithmic time (upper panel), and of linear time in the lower panel. The leftmost curve (in red) pertains to $T=1.0$ and in order to the right, the temperatures are $T= 0.80$, 0.7, 0.65, 0.6, 0.56, 054, 0.52, 0.50, 048,  0.46, 0.44 and 0.42. Note the extreme slowing down in the range $0.56<T<0.42$. The black (dashed) line at $T=0.6$ separates ``simple" time dependence at higher temperature, which can be well fitted to a stretched exponential function from ``non-trivial" time dependence at lower temperatures which is ill-fitted by any stretched exponential form. Our aim is to predict quantitatively the form and value of the relaxation function for all temperatures and times.}
\label{data}
\end{figure}

{\bf The relaxation function}. For the sake of this Letter we introduce a relaxation function that is made as follows. At time zero every large particle $i$ in the system is assigned a value $c_i=1$ and a neighbour list consisting of its $n$ (small or large) nearest neighbours. In time nearest neighbours wander into the yonder,  and when the $i$th particle loses one (respectively two, three)  of its nearest neighbours, having at that time  $n-1$ (respectively $n-2, n-3$) of the original list, it is assigned a value $c_i(t)= 2/3, 1/3$ and 0 respectively. The relaxation function that
we monitor is $C(t)\equiv (2/N)\sum_i^N c_i(t)$. Fig.~\ref{data} shows how the relaxation function
decays in logarithmic time for different temperatures as indicated in the figure legend. While at high temperatures ($T> 0.56$) the function can be fitted to a stretched exponential, for lower temperatures a long tail develops (starting at $T=0.56$), destroying any fit to a stretched exponential, as can be seen
in the lower panel where the same data is presented in linear time. Note the extreme slowing down
exhibited in the lowest eight temperatures.  

\begin{figure}
\centering
~\hskip -1.4 cm
\includegraphics[width=0.65\textwidth]{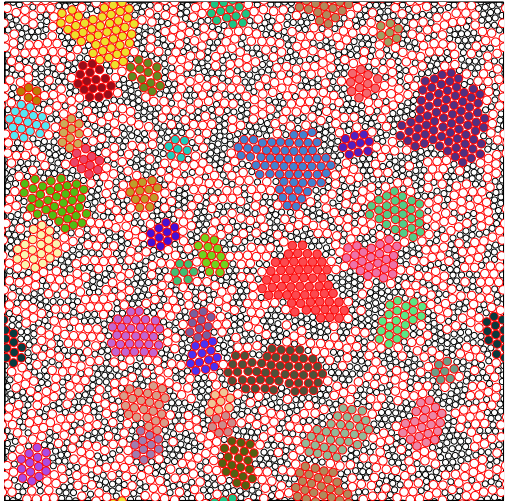}
\caption{(Color online). A snapshot of the system at $T=0.44$. In colours we highlight the clusters of large particles in  local hexagonal order. The colours have no meaning.}
\label{clusters}
\end{figure}

To gain a quick insight as to the nature of the slowing down it is advantageous to watch a movie
of the decay of the relaxation function. Such a movie is provided as a supplementary information
to this Letter \cite{movie}. Color coding in green every particle whose assigned value $c_i(t)>0$, it is obvious from the movie that a fraction of particles lose their green colour very rapidly, whereas the non-zero values of $C(t)$ at longer times are contributed entirely by {\em clusters} of large particles in 
local hexagonal order.  This visual observations brings us to the central theme of this Letter.
We propose that the present glass-former belongs to a generic class of glass-forming systems that have competing states (of crystalline order or not) which are locally close in energy to  the ground state (which is typically unique). Upon rapid cooling, such systems exhibit patches (or clusters) of these competing states which become locally stable in the sense of having a relatively high local shear modulus. It is in between these clusters where ageing, relaxation and plasticity under strain can take place. A still picture
of a typical realization of this system is provided in Fig. \ref{clusters} where the locally hexagonal
clusters of large particles are highlighted in colour (no meaning to different colours). For clarity
of presentation we do not show in this figure the local hexagonal patches of small particles, or any other
cluster of a competing phase.

\begin{figure}
\centering
~\hskip -0.8 cm
\includegraphics[width=0.55\textwidth]{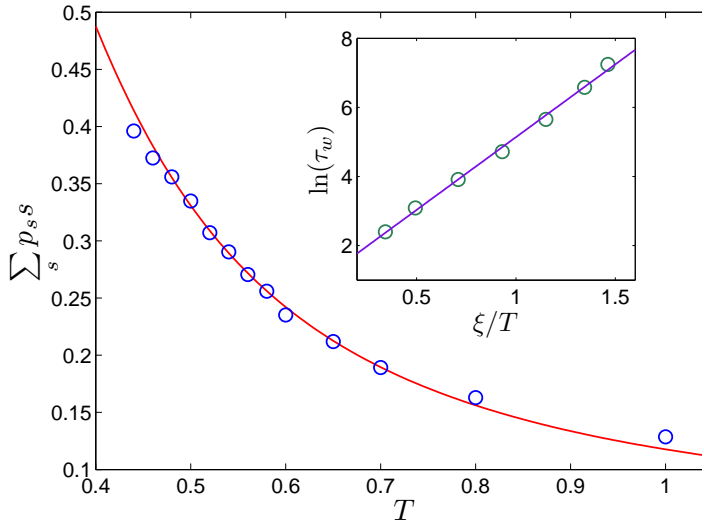}
\caption{(Color online). A test of Eq. \ref{estimate} with $\Delta =1.31$. In the inset we demonstrate Eq. (\ref{tauw}) with $\mu=4.212$. Here $\tau_w$ was measured directly as the decay time of the relaxation function of the whey Eq. (\ref{whey}), and $\xi(T)$ was computed from Eq. (\ref{xi}).}
\label{fit}
\end{figure}

{\bf Cluster decomposition formula}. In the rest of this Letter we show that this insight is the basis of a quantitative theory of the
relaxation function. To this aim we separate in our mind the clusters (here we only take into
account the clusters of large particles in hexagonal local order, which are like curds) from the rest of the system, which we refer to as the whey. In the whey the relaxation process is dominated by `defects', mainly large particles in pentagonal cages and small particles in heptagonal cages \cite{07ABHIMPS,07HIMPS}. The relaxation process necessitates a pair-wise annihilation of
such defects, because of the Euler constraint (requiring on the torus an average of six neighbours per
particle). If the average distance between such `defects' is a temperature dependent typical $\xi(T)$, the analysis of \cite{07ILLP,08EP} showed that the relaxation of such defects is determined by a typical relaxation time $\tau_w(T)$ which depends on the temperature according to
\begin{equation}
\tau_w(T) \propto e^{\mu\xi(T) /T} \ , \label{tauw}
\end{equation}
where $\mu$ is a constant having dimensions of $k_B$. Clearly, in order to estimate the typical distance between such defects we need to know how many
particles belong to clusters. Denote the number of clusters of size $s$ by $N_s$;  then $s p_s=s N_s/N$ is the probability to find a large particle in a cluster of size  $s$. The number density of large particles in the whey is $n_w=1-\sum_s p_s s$. To estimate $\xi$ we denote by $H_c$ and $H_w$ the average enthalpy of a large particle inside a cluster and in the whey respectively, and $\Delta=H_w-H_c$. There are  $g_w\approx (2^6 -1)/6+2^7/7$ ways to organize the neighbours of a large particle in the whey (neglecting the rare large particle in heptagonal neighbourhood), but only one way in the cluster. Thus the 
number density $n_c$ of particles in the clusters is
\begin{equation}
n_c \equiv \sum p_s s \approx  \frac{1}{1+g_w e^{-\Delta/T}} \ . \label{estimate}
\end{equation}
This estimate, with $\Delta=1.31$ agrees very well with the measured value of $n_c$, as can
be shown in Fig. \ref{fit}. From here we can determined $\xi(T)$, 
the typical distance between `defects', as
\begin{equation}
\xi\sim \sqrt{n_c}\ . \label{xi}
\end{equation}
To demonstrate the usefulness of this equation we have computed the relaxation function $C(t)$ only for particles in the whey, calling it $C_w(t)$, and found that it can be excellently fitted to the simple relaxation function
\begin{equation}
C_w(t) =  \alpha \exp\left[-\left(\frac{t}{\tau_w(T)}\right)^\beta\right]\ .
\label{whey}
\end{equation}
with $\alpha=0.95$ and $\beta=0.83$. The values of $\tau_w(T)$ were extracted from such measurements at different temperatures, and plotted as $\ln \left( \tau_w(T)\right)$ vs. $\xi(T)/T$. The results are shown as the inset in Fig. \ref{fit} in very satisfactory
agreement with Eq. (\ref{tauw}) with $\mu=4.212$. 
Note that it is highly non-trivial that exactly the same relaxation function describes the dynamics
in the whey at all temperatures. We will see that this changes dramatically when the clusters
are added to the picture. The coefficient of 0.95 represents an almost $T$-independent drop in $C_w(t)$
which results from thermal vibrations on the scale of one $\tau$. The exponent 0.83 is a fit that
appears to be specific to the present relaxation function.
\begin{figure}
\centering
~\hskip -0.8 cm
\includegraphics[width=0.55\textwidth]{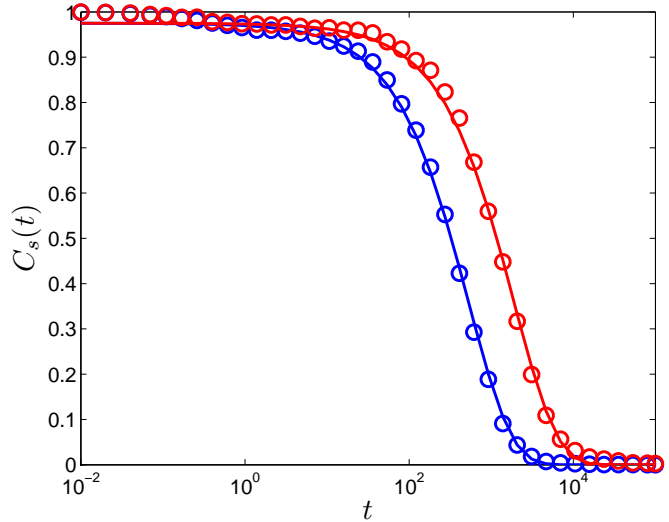}
\caption{(Color online). Demonstration that the relaxation of individual clusters of size $s$ (circles) follows
the same relaxation function as the whey (continuous line), but with a different relaxation time $\tau_s$. In blue we show $s=7$ at $T=0.5$ and in red $s=10$ at $T=0.46$. Note that 
we do not have similar fits for large clusters, and Eq. (\ref{taus}) is proposed as a model for
all $s$.}
\label{clusrelax}
\end{figure}

\begin{figure}
\centering
~\hskip -0.8 cm
\includegraphics[width=0.55\textwidth]{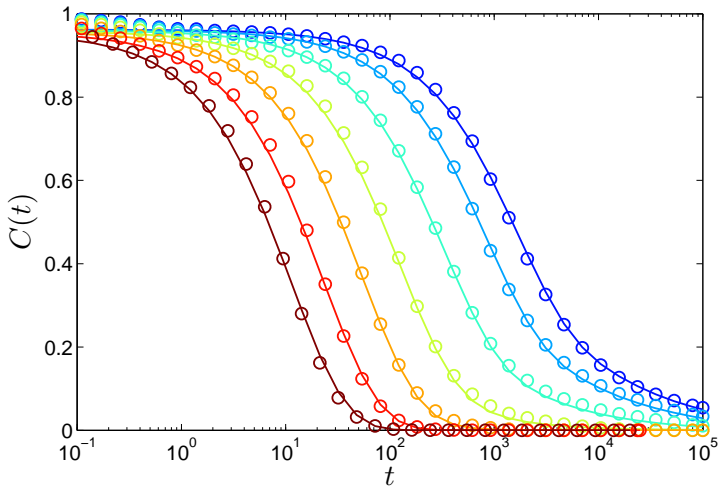}
\caption{(Color online). Comparison of the measured relaxation functions with the theoretical prediction
on the basis of the cluster decomposition formula (\ref{Ct}). From left to right the temperatures are
$T=$1, 0.8, 0.65, 0.56, 0.50, 0.46 and 0.44. Note the quantitative agreement
at all times and temperatures. The only parameter fitted here is $\sigma$ of Eq. (\ref{taus}).}
\label{final}
\end{figure}

Having understood the relaxation function of the whey we turn now to the relaxation function of the clusters. Our best statistics is for relatively small clusters, and we demonstrate in Fig. 
\ref{clusrelax} that the small clusters relax again {\em as the same simple function of time as the whey},
but with $\alpha\to \alpha'=0.98$ and an $s$ dependent relaxation time $\tau_s$ which can be fitted to
\begin{equation}
\tau_s = \tau_w \exp\left(\frac{\sigma s}{T}\right)=\exp\left(\frac{\mu \xi +\sigma s}{T}\right) \ . 
\label{taus}
\end{equation}
Note that this typical relaxation time for clusters of size $s$ goes smoothly to the whey when
$s\to 0$.  For larger clusters it is more difficult to say anything precise since the statistics deteriorates
very rapidly. Nevertheless, we will show now that the model (\ref{taus}) is sufficient for our purposes,
and that we can compute the total relaxation function quantitatively using this simple model. To do so
we present the relaxation function $C(t)$ as a sum over the whey and the clusters in the ``cluster decomposition formula (CDF)"
\begin{equation}
C(t) = n_w C_w(t)+
\sum_s p_s \alpha' \exp\left[-\left(\frac{t}{\tau_s(T)}\right)^\beta\right] \ . \label{Ct}
\end{equation}
It is important to note that at this point there is only one parameter left to fit, which is $\sigma$ in Eq. (\ref{taus}). The best fit is $\sigma=0.031$ with which we predict the relaxation functions as shown
in Fig. \ref{final}.
It is obvious that the CDF captures quantitatively both the time and the temperature dependence
of the relaxation function. We submit to the reader that a comparison of data and theory of this
quality had been quite rare in the subject of glass-forming systems.

We strongly believe that a similar approach should be relevant for a whole class of glass-forming
systems where clusters of competing phases can form upon rapid cooling \cite{GS82,93Cha,98KT,00TKV,06Lan,06ST,08HP}. It does not mean however
that one can automatically apply what had been done here to other cases. In each case the
physics of the glass-former should be carefully understood to identify what are the clusters that
dominate at longest times. For example in hydrogen bonded systems these may be compact clusters, fractal clusters  or chains of molecules, giving difficult to guess formulae for $\tau_s$ as a function
of the size $s$ of the cluster and of the temperature $T$. In addition, we should stress that in the
present case we have measured the distribution $p_s$; it is very desirable
to derive this distribution from statistical mechanical first principles, as well as to provide
a theoretical background to the fitted law (\ref{taus}). Notwithstanding these issues that remain
for future research we propose that this example provides unequivocal evidence that the existence
of locally favoured structure whose relaxation is much slower than the whey is fundamental to the understanding of the phenomenology of the glass-transition.


\end{document}